\documentclass[%
 reprint,
 amsmath,amssymb,
 aps,
prl,
]{revtex4-2}

\usepackage{xr}  

\usepackage{graphicx}
\usepackage{dcolumn}
\usepackage{bm}

\usepackage{hyperref}
\usepackage{xcolor}
\usepackage{amsthm}

\newcommand{\s}{{\rm S}}
\newcommand{\f}{{\rm F}}
\renewcommand{\r}{{\rm R}}
\newcommand{\h}{{\mathcal H}}
\newcommand{\rx}{{\mathbb R}}
\newcommand{\fl}[1]{\lfloor {#1}\rfloor }
\newcommand{\cx}{{\mathbb C}}
\newcommand{\hh}{{\mathfrak h}}
\newcommand{\bbbone}{\mathbf 1}

\newcommand{\z}{{\mathbf z}}
\newcommand{\w}{{\mathbf w}}
\newcommand{\upket}{{|\!\uparrow\rangle}}
\newcommand{\downket}{{|\!\downarrow\rangle}}
\newcommand{\upbra}{{\langle\uparrow\! |}}
\newcommand{\downbra}{{\langle \downarrow \!|}}

\newtheorem*{theorem}{Theorem}
\newtheorem{prop}{Proposition}

\begin{document}


\title{Bosonization of Noise Effects in Nonlocal Quantum Dynamics}

\author{Michele Fantechi}
\email{michele.fantechi@univ-lorraine.fr}
\affiliation{Université de Lorraine, CNRS, IECL, F-57000 Metz, France}

\author{\vspace*{-.3cm}Marco Merkli}
 \email{merkli@mun.ca}
\affiliation{Department of Mathematics and Statistics 
Memorial University of Newfoundland
St.~John's, NL, Canada A1C 5S7}

\date{\today}

\begin{abstract}
Quantum systems that interact non-locally with an environment are paradigms for exploring collective phenomena. They naturally emerge in various physical contexts involving long-range, many-body interactions. We consider a general class of such open systems characterized by a coupling to the environment which is inversely proportional to the square root of the environment size. We show that the induced system dynamics has a universal bosonic nature: the same evolution arises from coupling the system to a collection of noninteracting bosonic modes, independently of the microscopic structure of the original environment. This emergent ‘bosonization’ of the environment's influence results from the scaling of the coupling in the thermodynamic limit and is a manifestation of the quantum central limit theorem. While the effect has been observed in specific models before, we show that it is, in fact, a universal feature.
\end{abstract}

\maketitle

Collective phenomena in quantum many-body systems arise when many microscopic degrees of freedom interact. The dynamics cannot be reduced to the properties of individual particles, but instead it reflects long-range correlations, entanglement, and quantum coherence spread over many constituents. A classic example is superradiance in quantum optics, where a group of atoms interacts with light of wavelength much larger then the atom separation, causing the group to collectively emit a light pulse of much higher intensity than would be expected from independently emitting atoms \cite{D1,CGW,HL1,HL2,RSG,RDBE, SCLOCS,NLCZRT}. The loss of quantum coherence, called decoherence, between individual systems (qubits) through their interaction with a common noise source (environment, bath), is another collective phenomenon with wide-reaching consequences \cite{Schlosshauer, Zurek, Joosetal, MSB-prl}. 

Collective phenomena induced by long-range interactions are often described with {\it mean field theories} \cite{Spohn, AM, Mori, BCFN,BCFN2,BM,MR,CL,CoFaFaMe}, which are also used to derive effective non-linear evolution equations of complex quantum systems in the mathematical literature \cite{AGT,ESY1,ESY2,ESY3,FHSS,FK,Pickl}. A system interacting with $M$ components (particles, modes...) of a reservoir typically has an interaction energy of the order of $M$. In order to balance the interaction energy and the system energy as $M\rightarrow\infty$, mean field models consider a scaling of the coupling strength $g\propto 1/M$. This scaling suppresses all but the lowest (zero) order correlation effects of the reservoir on the  effective system dynamics | even though the latter depends on reservoir correlation functions of all orders. As a consequence, dynamical effects arising from fluctuations (bath correlations) can not be captured by the mean-field approximation. For instance, mean field theory predicts the relaxation to a stationary state in models of a quantum gas in an optical cavity | however, in reality non-stationarity is observed for long times, due to higher order correlation effects \cite{BJ}. Systems coupled indirectly via a common reservoir become entangled \cite{Braun} | however, entanglement in a multipartite system remains unaffected by a mean field type coupling to an environment \cite{FM,BJ}. In order to include effects caused by fluctuations, one can consider the scaling $g\propto 1/\sqrt M$ which unveils the result of first-order bath correlations on the system dynamics.  This scaling is said to be of {\it fluctuation}, or {\it mesoscopic} type. The terminology stems from the notion of {\it fluctuation observables}, which are extensive observables divided by the square root of the volume \cite{BCF14, BCF15, BCF16,BCFN17,BCFN,BCFN2,BCFNV, Verbeure}. A rigorous treatement of the flucutation regime is more demanding than the mean field one, because  correlation effects within the reservoir have to be taken into account. On a heuristic and numerical level, higher order cumulant expansions have also been considered for a class of central spin systems \cite{Fowler}.

In those systems
a single `central' spin interacts non-locally (`one-to-all') with a large number of `bath' or environmental spins, making up the reservoir $\r$ \cite{YAKE,BS1,BS2,KYLCG,KGIYLC,SCSDGKLG,CCL,HGA,DRC,Bernien,Asch}.  The mean field and fluctuation regimes of a central spin 1/2 coupled to bath spins 1/2 is analyzed in \cite{Carollo}.  It is shown that in the fluctuation regime ($g\propto 1/\sqrt M$) and $M\rightarrow \infty$, the effect of the reservoir can be encoded in a single bosonic mode communicating with the central spin, which results in an equivalent  Jaynes-Cummings model. The setup in \cite{Carollo} is  specific in several respects. Other than being two-level systems, $\s$ and $\r$ are coupled linearly via the collective bath ladder operators, conserving the total number of excitations. The reservoir dynamics is taken to be purely dissipative, without Hamiltonian component, purely driven by a pump and decay term between the two levels, and the reservoir is taken to be in the resulting equilibrium.

In this letter we explain that arbitrary quantum systems $\s$ mesoscopically coupled to reservoirs $\r$ of arbitrary $M\rightarrow\infty$ constituents, evolve as if they were linearly coupled to a set $\f$ of oscillators in their ground state. The equivalent bosonic fluctuation reservoir $\f$ is constructed in a simple, explicit, and physically transparent manner. We  show that for  generic $\s\r$ models where components of $\r$ have $Q\ge 2$ energy levels, $Q-1$ independent oscillators will arise in the equivalent fluctuation reservoir $\f$. We  show  that entanglement is created within bipartite systems $\s$ by the indirect mesoscopically scaled contact with reservoirs. Our results hold for arbitrary initial states of $\r$ which are stationary under the uncoupled, Hamiltonian reservoir dynamics. The fluctuation reservoir $\f$ depends explicitly on the initial state of $\r$.  We also identify the quantum central limit theorem as the universal mechanism underlying the equivalence $\r\leftrightarrow\f$. Our key observation is that the $\s\r$ coupling  operators behave as non-commutative, normally distributed random variables ($M\rightarrow\infty$) \cite{CH,GW,GV,CE,Skeide}. Bosonic systems, such as collections of oscillators in quasi-free (Gaussian) states, are likewise characterized by normally distributed correlation functions. By matching these correlations, we show that we can replace the original reservoir $\r$ by an equivalent set of harmonic oscillators $\f$. Our results are mathematically exact. 
\smallskip

\begin{figure}[h!]
\centering
\includegraphics[width=.53\textwidth]{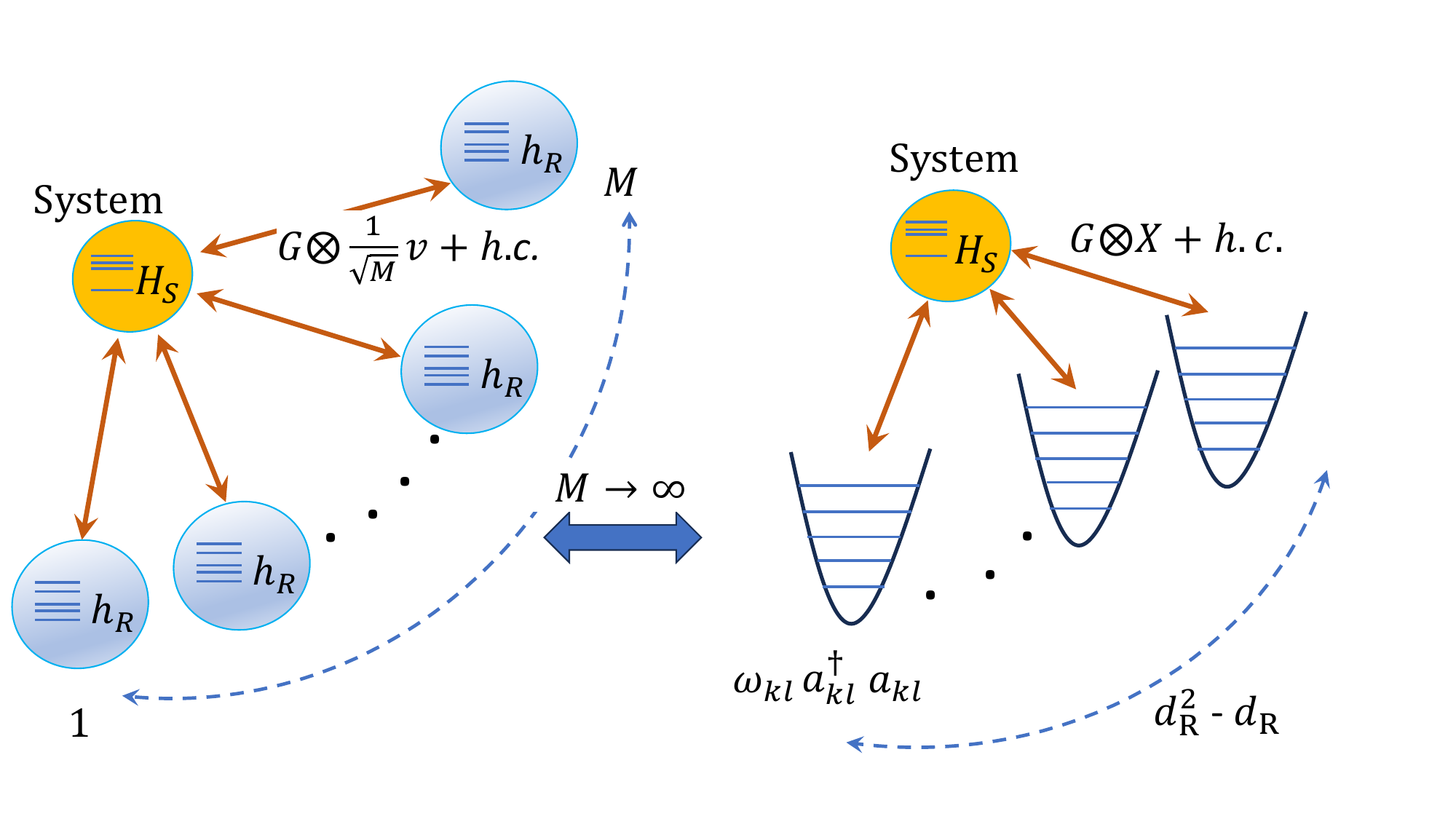}
\caption{{\bf Illustration of the main result.} A system $\s$ with Hamiltonian $H_\s$ interacting with two different reservoirs. {\it Left sketch}: $\s$ interacts equally with the $M$ elements of a reservoir, each a $d_\r$-level system with Hamiltonian $h_\r$. The interaction operator is  $\propto 1/\sqrt M$, called a mesoscopic scaling. {\it Right sketch}: $\s$ interacts with a reservoir of (maximally) $d_\r^2$ independent bosonic modes indexed by $(k,l)$, all in their ground state. Each oscillator corresponds to a transition $E_k\rightarrow E_l$ between energy levels of $h_\r$ allowed by the interaction $\s\r$ operator $v$. The interaction operator $X$ of $\s$ with the bosonic modes is linear in the creation and annihilation operators. The left-right arrow $\leftrightarrow$ indicates our {\it main result}: The reduced dynamics of $\s$ obtained from the left model (as $M\rightarrow\infty$) and the right model is the same. The result holds for arbitrary $\s$ and regardless of the fine details of the components of $\r$.}
\label{figure1} 
\end{figure}

\smallskip

{\it Model.} |  A quantum system $\s$ is coupled to a quantum reservoir $\r$ consisting of $M$ independent, identical components, see Figure \ref{figure1} (left part). The total complex is described by the Hilbert space
\begin{equation} \label{1}
\h_{\s\r,M}  = \h_\s\otimes\h_{\r,M},\quad 
\h_{\r,M}  = \h_\r\otimes\cdots \otimes\h_\r 
\end{equation}
($M$-fold product). We assume that $\dim\h_\s<\infty$ and $\dim\h_\r<\infty$. 
The dynamics is generated by the interacting hermitian Hamiltonian on $\h_{\s\r,M}$,
\begin{equation}
\label{3}
H = H_\s + H_{\r,M} +  V_M,
\end{equation}
where $H_\s$ is the (any) Hamiltonian of $\s$  and 
\begin{equation}
	\label{4}
H_{\r,M} = \sum_{m=1}^M h_\r^{[m]}
\end{equation}
is the sum of single-component Hamiltonians $h_\r^{[m]}$. Here, $h_\r$ is a fixed, arbitrary, hermitian operator and we use the notation 
\begin{equation}
\label{xm}
X_\r^{[m]} = \bbbone\otimes \cdots \otimes \bbbone\otimes X_\r \otimes\bbbone\otimes\cdots \otimes \bbbone,
\end{equation}
where the operator $X_\r$ sits on the $m$th factor in the $M$-fold product. The system-reservoir interaction operator in \eqref{3} is given by
\begin{equation}
\label{M}
V_M = \sum_{q=1}^Q G_q\otimes \frac{1}{\sqrt M} \sum_{m=1}^M v_q^{[m]} \ \  {\rm +\ \  h.c.},
\end{equation}
for some $Q\ge 1$, where $G_q$ and $v_q$ are (not necessarily hermitian) operators on $\h_\s$ and $\h_\r$, respectively.

We consider initial system-reservoir states (density matrices) of the form 
\begin{equation} \label{7.1}
\rho_{\s\r,M} = \rho_\s\otimes \rho_{\r,M}, \qquad  \rho_{\r,M} =\rho_\r\otimes\cdots\otimes\rho_\r,
\end{equation}
with an arbitrary system state $\rho_\s$ and where each component of $\r$ is in the same state $\rho_\r$, which is assumed to be stationary (for example a thermal equilibrium state), 
\begin{equation}
\label{stat}
e^{-it h_\r}\rho_\r e^{ith_\r} = \rho_\r.
\end{equation}
The interaction operators $v_q$ are normalized (shifted, centralized) to have vanishing expectation,
\begin{equation}
\label{centralized}
{\rm tr}_\r(\rho_\r v_q)=0.
\end{equation}
The object of interest is the reduced system density matrix in the thermodynamic limit,
\begin{equation}
\label{7-1}
\rho_{\s}(t)  = \lim_{M\rightarrow\infty} {\rm tr}_{\r,M} \Big( e^{-i t H_{\s\r,M}} \,\rho_{\s\r,M} \,e^{i t H_{\s\r,M}}\Big),
\end{equation}
where ${\rm tr}_{\r,M}$ stands for the partial trace over the reservoir degrees of freedom. 
\medskip

{\it Consequences of symmetry and scaling.} | We now discuss the mechanism and ideas which lead to our main result \eqref{026}. For ease of presentation we do this here for an interaction \eqref{M} with $Q=1$ and $G=G^\dag$, $v=v^\dag$. A mathematical proof of the general case is given in the supplemental material. The model has two important features:

(i) {\it Symmetry.} The full Hamiltonian \eqref{3} and the reservoir initial state \eqref{7.1} are symmetric under the exchange of reservoir components.

(ii) {\it $1/\sqrt M$ scaling.} The interaction \eqref{M} is scaled with this prefactor (termed {\it mesoscopic}).

The contribution of the reservoir to the dynamics is  encoded in the multi-time correlation functions
\begin{equation}
\label{sn1}
C_M(t_1,\ldots,t_n):= {\rm tr}_{\r,M}\big( \rho_{\r,M} \frac{1}{\sqrt M}S(t_1)\cdots\frac{1}{\sqrt M} S(t_n)\big),
\end{equation}
where $S(t)=\sum_{m=1}^Mv^{[m]}(t)$  (see \eqref{M}), with $v(t)=e^{ith_\r}ve^{-ith_\r}$. The product $S(t_1)\cdots S(t_n)$ in \eqref{sn1} is a multiple sum over indices $m_1,\ldots,m_n$. Each index $m$ indicates the reservoir element on which the operator $v^{[m]}$ acts (see \eqref{xm}). At first sight one might think that \eqref{sn1} would diverge as $M\rightarrow\infty$, because the multiple sum has $M^n$ terms and the scaling only provides a factor $M^{-n/2}$.  However, as ${\rm tr}(\rho_\r v(t))=0$ (see \eqref{stat} and \eqref{centralized}) and by the symmetry and product form of $\rho_{\r,M}$, \eqref{7.1}, terms in the multiple sum vanish whenever at least one $m$ differs from the others. So indices need to be paired up or appear in larger clusters for a nonzero contribution. A combinatorial argument shows that for  $M\rightarrow\infty$ all terms with larger clusters disappear as their number grows less than $O(M^{n/2})$. The scaling $M^{-n/2}$, with $M\rightarrow\infty$, selects precisely the terms where the indices $m_1,\ldots,m_n$ are paired, with each pair corresponding to the product of two operators $v(t)v(t')$ acting on the same reservoir “site” or component. This implies that the multi-correlation functions $C_M(t_1,\ldots,t_n)$ are expressed for $M\rightarrow\infty$, via the two-point function
\begin{equation}
\label{twopint}
c(t,t'):={\rm tr}_\r\big(\rho_\r v(t)v(t')\big)
\end{equation}
by Wick's theorem (see section ~\ref{sec:wickthm} of the supplemental material for a simple proof), just as in quantum field theory. Letting $\mathcal P_n$ denote the set of all pairings $\pi$, we have
\begin{align}
\lim_{M\rightarrow\infty} & C_M(t_1,\ldots,t_n) = \nonumber\\
&
\left\{
\begin{array}{cl}
\displaystyle \sum_{\pi\in\mathcal P_n} \prod_{j=1}^{n/2} c(t_{\pi(2j-1)},t_{\pi(2j)}) & \mbox{$n$ even}\\
\displaystyle 0 & \mbox{$n$ odd.}
\end{array}
\right. 
\label{sn2}
\end{align}
The formula \eqref{sn2} expresses the quantum central limit theorem. We prove it in Proposition \ref{prop2} in section ~\ref{sec:proofthm1} of the supplemental material.
\smallskip

{\it Equivalent bosonic (fluctuation) modes $\f$.} | We explain now how to build a collection of harmonic oscillators which reproduces the two-point function \eqref{twopint}. 
In the common eigenbasis $\{\chi_j\}$ of $\rho_\r$ and $h_\r$ we have
\begin{equation}
\label{6.2}
\rho_\r = \sum_{j=1}^{d_\r} p_j |\chi_j\rangle\langle\chi_j|,\quad h_\r = \sum_{j=1}^{d_\r} E_j |\chi_j\rangle\langle\chi_j|
\end{equation}
where the $0\le p_j\le 1$ are the populations of $\rho_\r$ and the $E_j\in \rx$ are the energies of $h_\r$ (possibly with $E_j=E_{j'}$ for different indices). Denoting the matrix elements of $v$ as $v_{kl}=\langle \chi_k| v|\chi_l\rangle$ the two-point function \eqref{twopint} reads,  
\begin{equation}
\label{match5.1}
c(t,t') = \sum_{k,l=1}^{d_\r} p_k |v_{kl}|^2 
 e^{-i(E_l-E_k)(t-t')}.
\end{equation}
We associate one harmonic oscillator $a^\dag_{kl}, a_{kl}$ to each pair $(k,l)\in\mathcal M':= \big\{ (k,l)\ \mbox{s.t.} \  p_k\neq 0,\ v_{kl}\neq 0 \big\}$, that is to each allowed ($v_{kl}\neq 0$) transition
$E_k\rightarrow E_l$ provided $E_k$ populated in $\rho_\r$. We take the frequency of the oscillators to be just the Bohr transition frequency, $\omega_{kl}=E_l-E_k$. This means physically that when the oscillators are coupled by an energy exchange interaction with $\s$, they can induce the same transitions in $\s$ as the original reservoir. The oscillator Hamiltonian is $H_\f = \sum_{(k,l)\in\mathcal M'} \omega_{kl} a^\dag_{kl} a_{kl}$. We introduce the field operator $\phi(t)=\sum_{(k,l)\in\mathcal M'} \sqrt{p_k}v_{kl} e^{i\omega_{kl}t} a^\dag_{kl} +{\rm h.c.}$ It is easily seen that its two-point function in the oscillator ground state $\rho_\f=|0\rangle\langle0|\otimes\cdots\otimes|0\rangle\langle 0|$ is precisely \eqref{match5.1}, ${\rm tr}\big(\rho_\f \phi(t)\phi(t')\big) = c(t,t')$. This implies that the new bath $\f$ consisting of all the oscillators, coupled to $\s$ linearly  via the interaction operator $G\otimes \phi(0)$, induces the same dynamics \eqref{7-1} in $\s$ as the original reservoir. One shows this by realizing that the Dyson series expansions of the dynamics of $\s$ resulting from the $\s\f$ interaction and from the $\s\r$ interaction are exactly the same. Indeed, these expansions express the system dynamics in terms of multi-correlation functions of the corresponding reservoirs ($\r$ or $\f$) and for both reservoirs, those multi-correlation functions are expressed in terms of the two-point functions via Wick's theorem (as in \eqref{sn2}). By design the two-point functions of $\r$ and $\f$ are the same, hence so are the multi-correlation functions and thus so are the Dyson series expansions. We present the technical details of the arguments above in sections \ref{sec:proofthm1} and \ref{sec:emerGS} of the supplemental material. Above is the outline of the result and the main ideas for $Q=1$, $G^\dag=G$ and $v^\dag=v$ in \eqref{M}. Now we present the general case.
\smallskip

{\it Main result.} | Consider the model \eqref{1}-\eqref{7-1}.  We introduce the `fluctuation reservoir' $\f$, which is made of independent harmonic oscillators $a^\dag_{kl}, a_{kl}$ in their ground state
\begin{equation}
\label{rhovac}
\rho_\f = |0\rangle\langle 0|\otimes\cdots \otimes |0\rangle\langle 0|.
\end{equation}
The oscillators $a_{kl}^\dag$, $a_{kl}$ are indexed by $(k,l)$ such that $E_k$ is an energy level occupied in $\rho_\r$ ($p_k\neq 0$) and at least one of the interaction operators $v_q$ in \eqref{M} allows the transition $E_k\rightarrow E_l$ or $E_l\rightarrow E_k$. In other words, the oscillators are indexed by $(k,l)$ in the set of modes,
\begin{align}
 \mathcal M  := \big\{ & (k,l) \mbox{\ s.t. $p_k\neq 0$ and} \label{30}\\
& \mbox{$[v_q]_{kl}$ or $[v_q]_{lk}\neq 0$ for some $q=1,\ldots, Q$}\big\}.
\nonumber
\end{align}
The oscillator $(k,l)$ has the frequency $\omega_{kl} =E_l-E_k$ and we define the Hamiltonian
\begin{equation*}
H_\f = \sum_{(k,l)\in \mathcal M} \omega_{kl}a^\dag_{kl}a_{kl}.
\end{equation*}
We couple the system $\s$ to $\f$  via the interacting Hamiltonian 
\begin{equation}
\label{MX}
H_{\s\f} = H_\s +H_\f + \sum_{q=1}^Q G_q\otimes X_q + G_q^\dag\otimes X_q^\dag,
\end{equation}
with $X_q$ defined as
\begin{equation}
\label{12}
X_q = \sum_{(k,l)\in\mathcal M} \sqrt{p_k} \big( [v_q]_{lk}\, a^\dag_{kl} + [v_q]_{kl}\, a_{kl}\big).
\end{equation}
The $[v_q]_{kl}=\langle\chi_k|v_q|\chi_l\rangle$ are the matrix elements of the interaction operator $v_q$ \footnote{For $Q=1$ and $G=G^\dag$, $v=v^\dag$ the Hamiltonian \eqref{MX} produces the situation described in the previous paragraph. Note also that $[v_q]_{lk} = \overline{([v_q]^\dag)_{kl}}$ (the bar denotes the complex conjugate).}. Our main result is:

\begin{theorem}
\label{thm1}
The system dynamics $\rho_\s(t)$ defined in \eqref{7-1}, is equivalently given, for all $t\in\rx$, by
\begin{equation}
\label{026}
\rho_\s(t) = {\rm tr}_\f \big(e^{-it H_{\s\f}} (\rho_\s\otimes\rho_\f) e^{it H_{\s\f}}\big)
\end{equation}
 where $H_{\s\f}$ is the interacting $\s\f$ Hamiltonian \eqref{MX} and $\rho_\f$ is the ground state \eqref{rhovac}. 
\end{theorem}

{\it Discussion and examples.} | 
The theorem shows that one can use a bath of oscillators in their ground state to model the dynamics of any system $\s$ within the class of $\s\r$ models \eqref{1}-\eqref{7-1} presented above. The form of the equivalent fluctuation reservoir $\f$ is not unique. $\f$ should be Gaussian (quasi-free) and have the same two-point function as $\r$. Intuitively, the reservoir $\f$ should provide the system $\s$ with the same set of transition energies $E_l-E_k$ and the same transition probabilities as $\r$ does, in order to drive the same system dynamics. A realization of $\f$ is the ground state oscillator bath \eqref{rhovac}-\eqref{12}, where the mode $a^\dag_{kl}$ enables the transition $E_k\rightarrow E_l$. We call this the standard representation of $\f$, as it is suitable for all models. In special situations (for parameter relations compatible with thermal transition probabilities), one may realize $\f$ as a collection of thermal modes at appropriate inverse temperatures $\beta_{kl}$, each availing $\s$ with transition energies $E_k-E_l$ and $E_l-E_k$ simultaneously. We explain this with more precision in the section \ref{sec:emerGS} of the supplemental material. The non-uniqueness of $\f$ is not surprising, as in general, different coupled system-reservoir dynamics can generate the same reduced system dynamics. This is akin to different multivariate probability distributions having a fixed marginal.

The number of oscillators in $\f$ is the cardinality of $\mathcal M$, \eqref{30}. It satisfies $|\mathcal M|\le d_\r\times{\rm rank}\, \rho_\r$, where $d_\r=\dim \h_\r$ and the rank of $\rho_\r$ is the number of nonzero populations $p_k$. If all transitions are allowed by the $v_q$ then $|\mathcal M|=d_\r \times {\rm rank}\, \rho_\r$. For `off-diagonal' interactions, $[v_q]_{kk}=0$ for all $q$ and $k$ (and all other transitions allowed), $\f$ has  $|\mathcal M|=(d_\r-1)\times {\rm rank}\,\rho_\r$ oscillators. A pure state $\rho_\r=|\psi_\r\rangle\langle\psi_\r|$ has rank one and the reservoir $\f$ has at most $d_\r$ oscillators. For a reservoir $\r$ made of spins, $d_\r=2$ we need $2$ oscillators in $\f$. If the interaction is off-diagonal, $[v_q]_{kk}=0$, then $\f$ consists of a single oscillator. The corresponding $\s\f$ model is then of the Jaynes-Cummings type as in \cite{Carollo}. On the other hand, thermal states $\rho_\r\propto e^{-\beta h_\r}$ ($\beta<\infty$) have full rank $d_\r$ leading to a maximal number $d_\r^2$ of oscillators (or $d_\r^2-d_\r$ for off-diagonal $v_q$).

{\it Qubit (spin) system.} Consider both $\s$ and each component of $\r$ to be spins $1/2$ and let $H_\s$ and $\rho_\s$ be any spin 1/2 Hamiltonian and density matrix. Take $h_\r=\omega_\r\upket\upbra$ with $\omega_\r>0$ and $\rho_\r = w\downket\downbra+(1-w)\upket\upbra $ with $0\le w\le 1$. For the interaction \eqref{M} take $Q=1$ and $G=\downket\upbra$, $v=\upket\downbra$ so that excitations are exchanged between $\s$ and $\r$. The reservoir populations and energy levels \eqref{6.2} in this example are, $p_1=1-w$, $p_2=w$, $E_1=0$, $E_2=\omega_\r$. If $w\neq 0,1$ (mixed state) then the set of modes \eqref{30} is $\mathcal M = \{(1,2), (2,1)\}$ and $\f$ consists of two oscillators $a^\dag, a$ and $b^\dag,b$ at frequencies $\omega_a=\omega_\r$ and $\omega_b=-\omega_\r$. The interaction operator \eqref{12} is $X=w^{1/2}\, a^\dag+(1-w)^{1/2}\, b$. In the case $w=0,1$ (pure state) $\mathcal M$ reduces to a single point and $\f$ consists of only one oscillator.  We introduce the new, effective modes $A=(1-2w)^{-1/2} (w^{1/2} a^\dagger + (1-w)^{1/2} b)$ and $B=(1-2w)^{-1/2} ((1-w)^{1/2} a + w^{1/2} b^\dagger)$ (for $0<w<1/2$) such that $[A,A^\dagger]=[B,B^\dagger]=\mathbf 1$, $[A,B]=[A,B^\dagger]=0$. 
In terms of these independent oscillators, the $\s\f$ Hamiltonian becomes
$$
H_{\s\f} = H_\s-\omega_\r (A^\dag A- B^\dag B) +G\otimes A+G^\dag\otimes A^\dag.
$$
Therefore, $\s$ is coupled to a single effective mode $A$ \footnote{We thank a referee to point this out to us.}.

{\it Emergence of multiple fluctuation modes.} Let $\s$ be arbitrary with any Hamiltonian $H_\s$ and any initial state $\rho_\s$. Take reservoir components to be $Q$-level systems ($Q\ge 2$), $h_\r=\sum_{q=1}^Q E_q |q\rangle\langle q|$ and $\rho_\r=|1\rangle\langle 1|$. Take a $\s\r$ coupling which induces transitions between $|1\rangle$ and $|q\rangle$, $q\ge 2$, namely \eqref{M} with arbitrary $G_q$ and $v_q= |q\rangle\langle 1|$. The mode set \eqref{30} is $\mathcal M=\{(1,q): q=2,\ldots,Q\}$ and we denote by $a_q\equiv a_{1,q}$ the corresponding bosonic operators of $\f$. The $\s\f$ Hamiltonian \eqref{MX} couples $\s$ to $Q-1$ independent modes of $\f$ with frequencies $\omega_q=E_q-E_1$,
$$
H_{\s\f} = H_\s +\sum_{q=2}^Q \omega_q a^\dag_q a_q + \sum_{q=2}^Q G_q\otimes a^\dag_q +{\rm h.c.}
$$

{\it Creation of intra-system entanglement.} The interaction of two systems $S_1$ and $S_2$ to a common environment $\r$ generally produces entanglement between $S_1$ and $S_2$, even if they are not coupled directly \cite{Braun}. This effect is not captured in mean-field models. It is shown in \cite{FM} that the effective propagator of multipartite systems $S=S_1+\cdots +S_N$ coupled in the mean field way to a common  reservoir, is the product of local unitaries (with time-dependent generators). The intra-system entanglement is thus constant in time and initially disentangled states of $\s$ stay so when interacting with $\r$. The following simple example shows that the creation of intra-system entanglement is restored by the interaction with a common bath in the fluctuation scaling. Take an arbitrary bipartite $S=S_1+S_2$ with any uncoupled $H_\s=H_{\s_1}+H_{\s_2}$ and a reservoir with two-level components, $h_\r=E_1|1\rangle\langle 1|+E_2|2\rangle\langle 2|$, say in the ground state $\rho_\r = |1\rangle\langle 1|$. Take the coupling \eqref{M} with $Q=1$,  $G=\Gamma_1+\Gamma_2$ (any coupling operators of $S_1$ and $S_2$) and $v=|1\rangle\langle 2|$. The fluctuation reservoir $\f$ has a single mode $a$ and the $\s\f$ Hamiltonian \eqref{MX} is ($\omega_\f=E_2-E_1$),
\begin{equation}
\label{HBraun}
H_{\s\f} = H_{\s_1}+H_{\s_2}+ \omega_\f a^\dag a + (\Gamma_1+\Gamma_2)\otimes (a^\dag +a).
\end{equation}
This Hamiltonian generates entanglement between $S_1$ and $S_2$ \footnote{For $S_1$ and $S_2$ each a qubit the Hamiltonian \eqref{HBraun} is that of  \cite{Braun} but with a single reservoir mode. Using the standard PPT criterion one can readily check that entanglement between the qubits is created by the indirect contact with the single mode}.

\smallskip

{\em Non-demolition models.} Consider $Q=1$ and commuting $H_\s$ and $G$ (\eqref{3}, \eqref{M}), $
H_\s=\sum_{j=1}^N e_j|\psi_j\rangle\langle\psi_j|$, $G=\sum_{j=1}^N g_j|\psi_j\rangle\langle\psi_j|$,
where $\{\psi_j\}_{j=1}^N$ is an orthonormal basis of $\h_\s=\mathbb C^N$ and $e_j\in\mathbb R$, $g_j\in\mathbb C$ ($G$ is not necessarily hermitian). Let $h_\r$,  $\rho_\r$ and $v$ be arbitrary, fixed. As $H_\s$ commutes with $H_{\s\f}$ the populations of $\s$ (diagonal density matrix elements) in the dynamics $\rho_\s(t)$, \eqref{026}, are independent of time. The coherences (off-diagonals) evolve independently, $\langle\psi_m|\rho_\s(t)|\psi_n\rangle = e^{-it(e_m-e_n)} D_{m,n}(t) \langle\psi_m|\rho_\s(0)|\psi_n\rangle$, where the decoherence functions is,
\begin{equation}
D_{m,n}(t) =\prod_{(k,l)\,:\, k\neq l} D_{m,n}(k,l,t).
\label{tDq}
\end{equation}
Consider the case $g_n\in\mathbb R$ and $v=v^\dagger$. Then the modulus of \eqref{tDq} is (see End Matter below)
\begin{equation}
\begin{split}
|&D_{m,n}(t)| \label{os46}\\
&= \exp\big[-8(g_m-g_n)^2\!\!\sum_{(k,l)\, :\, k\neq l}  p_k |v_{kl}|^2   \tfrac{\sin^2(t \omega_{kl}/2)}{\omega_{kl}^2}\big].
\nonumber
\end{split}
\end{equation}
The system undergoes a quasi-periodic evolution in time, in contrast to what happens for open systems coupled to reservoirs having a continuum of modes, such as the spin boson model. Those systems show decoherence, that is $D_{m,n}(t)\rightarrow0$ as $t\rightarrow\infty$. Nevertheless, for times $t<\!\!<\min_{k\neq l}\frac{1}{|\omega_{kl}|}$
we have the onset of decoherence, 
\begin{align}
\begin{split}
\big| D_{m,n}(t)\big| & \sim  \exp\Big[-2(g_m-g_n)^2 t^2\sum_{k,l=1}^{d_\r} |v_{kl}|^2 p_k\Big]\\
& = e^{-t^2(2{d_\r}-1)u^2 },
\end{split}
\label{os47}
\end{align}
where in the equality we took a homogeneous coupling, 
$
|v_{kl}|=\left\{
\begin{array}{cl}
 u & k\neq l\\
 0 & k=l
\end{array}
\right.
$. 
The decoherence rate \eqref{os47} is largely model-independent. It does not depend on the initial state $\rho_\r$ nor on the eigenvalues $E_j$ of $h_\r$. \eqref{os47} stays valid for longer times if the energy gap narrows, which is typically the case for increasing $d_\r$.  

{\it Summary.} | We study quantum systems $\s$ coupled to  reservoirs $\r$ of size $M$ through non-local interactions, scaled mesoscopically, proportional to $1/\sqrt{M}$, in the limit $M\to\infty$. We show that the reduced density matrix of the system $\s$ evolves exactly as if it were interacting with a different reservoir $\f$, consisting of independent bosonic modes initially in their ground states, and coupled through a simple, explicit, linear interaction. Our findings reveal that free bosonic reservoirs universally emerge from mesoscopically scaled, non-local couplings. We plan to use this framework to analyze the interplay of mean field and fluctuation effects in systems far from equilibrium (transport) and describe thermodynamic effects at arbitrary coupling strengths.

\smallskip

{\bf Acknowledgements.} M.M. was supported by a Discovery Grant from NSERC, the Natural Sciences and  Engineering Research Council of Canada. This research was funded, in whole or in part, by l’Agence Nationale de la Recherche (ANR), project ANR-22-CE92-0013. We are grateful to three referees for evaluating this work and giving us constructive feedback which improved the paper. We also thank the Associate Editor for their expert handling of the manuscript. 

\bibliographystyle{apsrev4-2}
\bibliography{ArXiv-bibliography.bib}

\section*{End matter}
\label{sec:endmatter}

The product structure in \eqref{tDq} comes from the fact that $\rho_\f$ is factorized and that all modes $(k,l)$ are independent. Each mode has its individual decoherence function, $D_{m,n}(k,l,t) = \langle 0| Y_{kl}(t,n) Y_{kl}(-t,m)  |0\rangle$, where $Y_{kl}(t,n) = e^{it( \omega_{kl} a^\dag a +\sqrt{p_k} \phi(\overline g_n \overline{v_{kl}} +g_n v_{lk})}$ and where $|0\rangle$ is the vacuum of the single mode with creation and annihilation operator $a^\dag,a$ and $\phi(x)=xa^\dag + \overline x a$ (for $x\in\mathbb C$). A standard polaron transformation type calculation gives
(see for instance Lemma 1 of \cite{MarMer24}), $D_{m,n}(k,l,t) =  e^{i A} \big\langle 0\big| W\Big(\frac{e^{i \omega_{kl} t}-1}{i \omega_{kl}}\sqrt{2p_k} 
\big\{ \big(\overline g_n-\overline g_m)\overline{v_{kl}} + (g_n-g_m)v_{lk}\big\}\Big)\big|0
\big\rangle$ for some phase $A$ (which has an explicit albeit a bit cumbersome expression, depending on $m,n,k,l,t$) and where $W(x)=e^{i\phi(x)}$ is the Weyl operator. The average of $W(x)$ in the state $|0\rangle$ is the Gaussian $\langle 0| W(x)|0\rangle =e^{-\frac14|x|^2}$, so we obtain $|D_{m,n}(k,l,t)| 
 = \exp\big[-2 p_k \frac{\sin^2(t \omega_{kl}/2)}{\omega_{kl}^2}  \big| \big(\overline g_n-\overline g_m)\overline{v_{kl}}
+ (g_n-g_m)v_{lk}\big|^2\big]
$. 
For the case $g_n\in\mathbb R$ and $v=v^\dagger$ we have $|(\overline g_n-\overline g_m)\overline{v_{kl}} + (g_n-g_m)v_{lk}|^2= 4(g_m-g_n)^2 |v_{kl}|^2$ and the expression in the main text follows.

\newpage

\onecolumngrid

\begin{flushleft}
    \Large \textbf{Supplemental Material} \normalsize - \textbf{Bosonization of Noise Effects in Nonlocal Quantum Dynamics.}
\end{flushleft}

\normalsize 

\renewcommand{\thesection}{\Alph{section}}
\renewcommand{\thesubsection}{\thesection.\arabic{subsection}}
\setcounter{section}{19}

\setcounter{section}{18}

\section{Proof of Theorem}
\label{sec:proofthm1}

\renewcommand{\theequation}{\Alph{section}.\arabic{equation}}

In the sequel, by ``the theorem'', we shall mean the main result of the letter | stated as a theorem involving \eqref{026}. To show the theorem we first write the dynamics $\rho_\s(t)$, \eqref{7-1}, in Dyson series form (Proposition \ref{prop1}). The limit $M\rightarrow\infty$ brings about the bosonic nature of the coupling operators,  which are of fluctuation type $\frac{1}{\sqrt M}\sum_{m=1}^M v_q^{[m]}$. This results in reservoir partial trace terms having a structure as dictated by the Wick theorem (Proposition \ref{prop2}). We then devise a fluctuation reservoir in a Gaussian state of independent bosonic modes, resulting in the same Wick structure. In what follows it is sometimes practical to use the following equivalent notion for a state, given either by a density matrix $\rho$ or a linear functional $\omega$ (acting on observables):
$$
\rho \leftrightarrow \omega(\cdot) = {\rm tr}(\rho\, \cdot\,).
$$

\subsection{Dynamics as $M\rightarrow\infty$ by the Dyson series}

We decompose the Hamiltonian \eqref{3} as
\begin{equation}
\label{13.1}
H_M=H_M^0 + V_M, \mbox{\quad with\quad} H_M^0 = H_\s + H_{\r,M},
\end{equation}
see also \eqref{M}. The Dyson series expansion gives
\begin{eqnarray}
\lefteqn{
{\rm tr}_{\r,M} \big( e^{it H_M^0} e^{-it H_{\s\r,M}} \rho_{\s\r,M}\, e^{it H_{\s\r,M}}e^{-it H_M^0}\big)}\nonumber\\
&=& \sum_{n\ge 0}(-i)^n \int_{0\le t_n\le\cdots \le t_1\le t} dt_1\cdots dt_n \ {\rm tr}_{\r,M}\big( 
\big[V_M(t_1), [ \cdots [V_M(t_n),\rho_\s\otimes\rho_{\r,M}]\ldots ]\big]\big),
\label{13-1}
\end{eqnarray}
where
\begin{equation}
\label{M'}
V_M(t) = \sum_{q=1}^Q G_q (t)\otimes V_{\r,M,q}(t)\ +{\rm h.c.},\qquad V_{\r,M,q}(t)=\frac{1}{\sqrt M} \sum_{m=1}^M v_q(t)^{[m]},
\end{equation}
with
\begin{equation}
G_q(t) =e^{itH_\s}G_q e^{-it H_\s},\qquad v_q(t)=e^{it h_\r}v_q e^{-it h_\r}.
\end{equation}
The series \eqref{13-1} converges in any norm on the bounded operators on $\h_\s$, for all values of $t\in\rx$ and $M\in\mathbb N$.

\begin{prop}
\label{prop1}
Recall the definition \eqref{7-1} of $\rho_\s(t)$. We have
\begin{align}
e^{it H_\s} \rho_\s(t) &e^{-itH_\s} \nonumber \\
=&
\sum_{n\ge 0}(-i)^n \int_{0\le t_n\le\cdots \le t_1\le t} dt_1\cdots dt_n \lim_{M\rightarrow\infty} {\rm tr}_{\r,M}\big( 
\big[V_M(t_1), [ \cdots [V_M(t_n),\rho_\s\otimes\rho_{\r,M}]\ldots ]\big]\big).
\label{35.1}
\end{align}
\end{prop}

\medskip

{\it Proof of Proposition \ref{prop1}. } The task is to show that the series in \eqref{13-1} converges (absolutely) {\it uniformly} in $M$, so that the limit and the summation can be interchanged. To show uniform convergence we proceed as follows. Using the relation \eqref{M'} in \eqref{13-1} results in
\begin{align}
{\rm tr}_{\r,M} &\big( e^{it H_M^0} e^{-it H_{\s\r,M}} \rho_{\s\r,M}\, e^{it H_{\s\r,M}}e^{-it H_M^0}\big)  =\sum_{n\ge 0}(-i)^n \sum_{q_1,\ldots,q_n} \sum_{\sigma_1,\ldots,\sigma_n}\int_{0\le t_n\le\cdots \le t_1\le t} dt_1\cdots dt_n \nonumber\\
& \times {\rm tr}_{\r,M}\big( 
\big[G_{q_1}^{\sigma_1}(t_1)\otimes V^{\sigma_1}_{\r,M,q_1}(t_1), [ \cdots [G_{q_n}^{\sigma_n}(t_n)V_{\r,M,q_n}^{\sigma_n}(t_n),\rho_\s\otimes\rho_{\r,M}]\ldots ]\big]\big).
\label{13-1-1}
\end{align}
Here, we have introduced the variables $\sigma\in\{\pm1\}$ and we set for an operator $\mathcal O$
$$
\mathcal O^\sigma = \left\{
\begin{array}{ll}
\mathcal O & \mbox{for $\sigma=1$}\\
\mathcal O^\dag & \mbox{for $\sigma=-1$}.
\end{array}
\right.
$$
Consider fixed values of $q_1,\ldots,q_n$ and $\sigma_1,\ldots,\sigma_n$. Expanding the multiple commutator in \eqref{13-1-1} inside the trace yields $2^n$ terms, each of the form
\begin{align}
G_{q_{\ell_1}}^{\sigma_{\ell_1}}(t_{\ell_1})\cdots & \, G_{q_{\ell_L}}^{\sigma_{\ell_L}}({t_{\ell_L}}) \ \rho_\s \ G_{q_{\ell_{L+1}}}^{\sigma_{\ell_{L+1}}}(t_{\ell_L+1})\cdots G_{q_{\ell_n}}^{\sigma_{\ell_n}}(t_{\ell_n})\nonumber\\
&\otimes V^{\sigma_{\ell_1}}_{\r,M,q_{\ell_1}}(t_{\ell_1})\cdots V^{\sigma_{\ell_L}}_{\r,M,q_{\ell_L}}(t_{\ell_L})\ \rho_{\r,M}\  V^{\sigma_{\ell_{L+1}}}_{\r,M, q_{\ell_{L+1}}}(t_{\ell_L+1})\cdots V^{\sigma_{\ell_n}}_{\r,M,q_{\ell_n}}(t_{\ell_n})
\label{16.1}
\end{align}
for some $L$ and some permutation $j\mapsto \ell_j$ of the $n$ indices. We now analyze the trace over the reservoir factor of \eqref{16.1},
\begin{align}
& {\rm tr}_{\r,M}\Big(V^{\sigma_{\ell_1}}_{\r,M,q_{\ell_1}}(t_{\ell_1})\cdots V^{\sigma_{\ell_L}}_{\r,M,q_{\ell_L}}(t_{\ell_L})\ \rho_{\r,M}\  V^{\sigma_{\ell_{L+1}}}_{\r,M, q_{\ell_{L+1}}}(t_{\ell_L+1})\cdots V^{\sigma_{\ell_n}}_{\r,M,q_{\ell_n}}(t_{\ell_n})\Big)\nonumber\\
& = {\rm tr}_{\r,M}\Big(\rho_{\r,M} V_{r_1}^{\tau_1}(s_1)\dots V_{r_n}^{\tau_n}(s_n)\Big)\nonumber\\
&= \frac{1}{M^{n/2}}\sum_{m_1,\ldots,m_n=1}^M {\rm tr}_{\r,M}\Big(\rho_{\r,M}\,  v_{r_1}^{\tau_1}(s_1)^{[m_1]} \cdots v_{r_n}^{\tau_n}(s_n)^{[m_n]}\Big),
\label{35}
\end{align}
where in the first step we used the cyclicity of the trace, the $s$ are a permutation of the $t$, the $\tau$ are a permutation of the $\sigma$ and the $r$ are a permutation of the $q$. Below we show the bound 
\begin{equation}
B(M)\equiv \frac{1}{M^{n/2}}\  \Big| \sum_{m_1,\ldots,m_n=1}^M {\rm tr}_{\r,M}\Big(\rho_{\r,M}\,  v_{r_1}^{\tau_1}(s_1)^{[m_1]} \cdots v_{r_n}^{\tau_n}(s_n)^{[m_n]}\Big)\Big| \le \big(e \max_r\|v_r\|\big)^n n^{n/2},
\label{35--1}
\end{equation}
for any $n\ge 1$ and any $M\ge 1$. Given the bound \eqref{35--1} we estimate the Dyson series \eqref{13-1} in operator norm (on bounded operators acting on $\h_\s$) from above by the series
\begin{equation}
\label{theseries}
\sum_{n\ge 0} 2^n Q^n\big(2 e \max_q\|G_q\|\, 
\max_q\|v_q\| \, t\big)^n \ \frac{n^{n/2}}{n!},
\end{equation}
where the factor $2^n Q^n$ accounts for the multiplicity of the sums over $\sigma$ and $q$ in \eqref{13-1-1}. The series  \eqref{theseries} converges for all values of $\max_q\|G_q\|$, $\max_q\|v_q\|$ and $t$  (use for example the quotient test). Consequently, the Dyson series \eqref{13-1} converges uniformly in $M$ and we can interchange the limit and summation, so that \eqref{7-1}, \eqref{13.1}, \eqref{13-1} yield \eqref{35.1}. 

This concludes the proof of Proposition \ref{prop1} modulo a proof of \eqref{35--1}, which we give now. We distinguish two cases, $M\le n$ and $M>n$. For $M\le n$ we have $B(M)\le M^{-n/2} (\max_q\|v_q\|)^n M^n \le (\max_q\|v_q\|)^n n^{n/2}$, so that \eqref{35--1} holds. Let us now treat the case $M>n$.  It is convenient to make a change of variables in the sum in \eqref{35--1}. For  a given configuration $(m_1,\ldots,m_n)\in \{1,\ldots,M\}^n$, there is a unique number $p_1\in\{0,\ldots,n\}$ of indices in the configuration equal to $1$, and a unique number $p_2\in\{0,\ldots,n\}$ of indices equal to $2$, and so on, and there is a unique number  $p_M\in\{0,\ldots,n\}$ of indices in the configuration which are equal to $M$. We have  $p_1+\cdots+p_M=n$. Conversely, given $(p_1,\ldots,p_M)\in \{0,\ldots,n\}^M$ such that $p_1+\cdots+p_M=n$ there are 
\begin{equation}
\label{multinomial}
{n\choose p_1,\ldots,p_M}=\frac{n!}{(p_1)!\cdots (p_M)!}
\end{equation}
different associated configurations $(m_1,\ldots,m_n)\in\{1,\ldots,M\}^n$ with such multiplicities $p_j$ (number of terms). The following identities guarantee that the multiplicities (number of terms) are accounted for correctly, 
\begin{equation}
\label{38}
\sum_{m_1,\ldots,m_n=1}^M 1 \ = M^n = \sum_{p_1,\ldots,p_M\ge 0\atop p_1+\cdots +p_M=n} {n\choose p_1,\ldots,p_M}.
\end{equation}
Not all the terms in the sum over the $p_j$ contribute to the sum \eqref{35--1}. Indeed, if any of the $p_j$ equals one, then the summand vanishes due to ${\rm tr}_\r(\rho_\r v_q^\sigma(s))=0$ (see \eqref{stat} and \eqref{centralized}). Hence all the $p_j$ which contribute to the sum are either zero or at least equal to two. As $p_1+\cdots + p_M=n$ this means that at most $\fl{n/2}$ of the $p_j$ can be nonzero. Therefore,
\begin{equation}
\label{41}
B(M) \le \frac{\max_q\|v_q\|^n}{M^{n/2}} \sum_{p_1,\ldots,p_M\ge 0 \atop p_1+\cdots+p_M=n} {n \choose p_1,\ldots,p_M} \chi(p_1,\ldots,p_M),
\end{equation}
where the function $\chi$ selects the configurations $(p_1,\ldots,p_M)$ with at most $\fl{n/2}$ nonzero $p_j$ (which are all $\ge 2$). Due to the symmetry of the sum in the $p_j$, we have 
\begin{equation}
\label{42}
\sum_{p_1,\ldots,p_M\ge 0 \atop p_1+\cdots+p_M=n} {n \choose p_1,\ldots,p_M} \chi(p_1,\ldots,p_M) = \sum_{\nu=1}^{\fl{n/2}} {M\choose \nu} \sum_{r_1,\ldots,r_\nu\ge 2 \atop r_1+\cdots+r_\nu=n} {n \choose r_1,\ldots,r_\nu} 
\end{equation}
where $\nu$ counts the number of nonzero $p_j$ | which are denoted $r_k$ | and ${M\choose \nu}$ counts the ways we can choose which of the $p_j$ are nonzero. Since $M>n$, Pascal's triangle tells us that the binomial coefficient is maximal for $\nu=\fl{n/2}$, so
\begin{equation}
\label{43}
{M\choose \nu}\le {M\choose \fl{n/2}} = \frac{M!}{(M-\fl{n/2})! \fl{n/2}!} \le \frac{M^{\fl{n/2}}}{\fl{n/2}!}.
\end{equation}
We combine \eqref{41}-\eqref{43} to obtain,
\begin{equation}
\label{44}
B(M)\le \frac{(\max_q\|v_q\|)^n}{\fl{n/2}!} \sum_{\nu=1}^{\fl{n/2}} \nu^n \le (\max_q\|v_q\|)^n\frac{\big(\fl{n/2}\big)^{n+1}}{\fl{n/2}!}\le (e\max_q\|v_q\|)^n \big(\fl{n/2}\big)^{n-\fl{n/2}+1/2}.
\end{equation}
In the first inequality, we estimated the sum over the $r_j$ in \eqref{42} above by $\nu^n$ using \eqref{38} and in the last step we used Stirling's bound, $\fl{n/2}!\ge \sqrt{2\pi \fl{n/2}} (\frac{\fl{n/2}}{e})^{\fl{n/2}}$. Finally, using the bound $\big(\fl{n/2}\big)^{n-\fl{n/2}+1/2}\le (n/2)^{\frac n2+1}\le n^{n/2}$ in \eqref{44} shows \eqref{35--1} for $M>n$.
This concludes the proof of Proposition \ref{prop1}.\hfill $\square$ 
\bigskip

Next we analyze the limit $M\rightarrow\infty$.

\begin{prop}
\label{prop2}
Let $w_j$, $j=1,\ldots,n$, be operators on $\h_\r$ such that $\omega_\r(w_j) \equiv {\rm tr}_\r(\rho_\r w_j)=0$ and set
$$
W_j = \frac{1}{\sqrt M}\sum_{m=1}^M w_j^{[m]}.
$$ 
Then we have 
\begin{equation}
\lim_{M\rightarrow\infty}{\rm tr}_{\r,M}\Big(\rho_{\r,M}\,  W_1 \cdots W_n\Big) =
\left\{
\begin{array}{cl}
\displaystyle \sum_{\pi\in\mathcal P_n} \prod_{j=1}^{n/2} \omega_\r\Big( w_{\pi(2j-1)} w_{\pi(2j)}\Big) & \mbox{for $n$ even}\\
\displaystyle 0 & \mbox{for $n$ odd}
\end{array}
\right. 
\label{45}
\end{equation}
where  $\mathcal P_n$ is the set of all permutations $\pi$ of $\{1,\ldots,n\}$ such that 
\begin{equation}
\label{perm'}
1=\pi(1)<\pi(3)<\cdots<\pi(n-1),\qquad \mbox{and}\qquad \pi(2j-1)<\pi(2j),\quad \mbox{for all $j$.}
\end{equation}
In particular, \eqref{45} holds for  $w_j=v_{r_j}^{\tau_j}(s_j)$, where $\tau_j\in\{\pm1\}$, $r_j\in\{1,\ldots,Q\}$ and $s_j\in\mathbb R$ are arbitrary.
\end{prop}
\medskip

{\it Proof of Proposition \ref{prop2}. } Not all configurations of the $(m_1,\ldots,m_n)$ contribute to the sum on the left side of \eqref{45}. If any of the indices $m_j$ is distinct from all the others, then the trace vanishes, due to the fact that $\rho_{\r,M}=\rho_\r\otimes\cdots\otimes\rho_\r$ and ${\rm tr}_\r(\rho_\r w_j)=0$, {\it c.f.~}\eqref{7.1}, \eqref{xm}. Therefore, in the multiple sum over the $m_j$, only the configurations count in which for some $2\le J\le n$, there are $k_2$ pairs of indices taking the same value ($1,\ldots,M$) within each pair but different values for different pairs, and there are $k_3$ triplets taking the same value within each triplet (but different values between triplets, and none equal to the values taken by the pairs), and so on, and there is a number $k_J$ of $J$-tuples which take the same value within each $J$-tuple but different values for different $J$-tuples (none of which are taken by any other of the previous $j$-tuples). Those numbers must add up to the total amount of indices,
\begin{equation}
\label{36.1}
\sum_{j=2}^J j k_j=n.
\end{equation}
Given $J$ and $k_1,\ldots,k_J$, each of the $k_2$ pairs takes a different value for the index $m=1,\ldots,M$, leading to $M(M-1)\cdots(M-k_2+1)\le M^{k_2}$ possibilities. Each of the $k_3$ triples have to take a different value for $m$ among the remaining $M-2k_2$ values. Hence $(M-2k_2)\cdots (M-2k_2-k_3+1)\le M^{k_3}$ possibilities. A similar bound holds for each of the groups of $j$-tuples. We thus obtain the bound
\begin{equation}
\Big| \frac{1}{M^{n/2}}\sum_{m_1,\ldots,m_n=1}^M {\rm tr}_{\r,M}\Big(\rho_{\r,M}\,  w_1^{[m_1]} \cdots w_n^{[m_n]}\Big)\Big| \le (\max_j\|w_j\|)^n \sum_{J=2}^n \sum_{k_2,\ldots,k_J\ge 0 \atop \sum_{j=2}^J j k_j=n} \frac{M^{k_2+\cdots+k_J}}{M^{n/2}}.
\label{35-1}
\end{equation}
Next, \eqref{36.1} gives $n = 2(k_2+k_3+\cdots +k_J) + n_0$ with $n_0=k_3+2k_4+3k_5\cdots +(J-2)k_J\ge 0$
and so 
$$
M^{k_2+\cdots+k_J} = M^{\frac n2-\frac{n_0}{2}}.
$$
We have $n_0\ge 1$ if $J>2$ and $n_0=0$ for $J=2$. Using this in \eqref{35-1} we see that the limit $M\rightarrow \infty$ vanishes unless $J=2$. Therefore the only terms surviving the $M\rightarrow\infty$ limit in the term on the left side of \eqref{45} are those for which the indices $m_1,\ldots,m_n$ take pairwise equal values and different values among pairs. 

To list all these terms we proceed as follows. For each such term, $m_1=m_{\pi(1)}$ equals some other index $m_{\pi(2)}$ with $\pi(1)<\pi(2)$. The next `free' $m$-index in the product $w_1^{[m_1]}\cdots w_n^{[m_n]}$ (not equal to $m_{\pi(1)}$ nor $m_{\pi(2)}$) is $m_{\pi(3)}$ for some $\pi(1)<\pi(3)$. That one is paired up with an $m_{\pi(4)}$ with $\pi(3)<\pi(4)$. Continuing this way we obtain the pairing
\begin{equation}
\label{47}
\big(w_{\pi(1)}, w_{\pi(2)}\big), \big(w_{\pi(3)}, w_{\pi(4)}\big),\ldots, \big(w_{\pi(n-1)}, w_{\pi(n)}\big).
\end{equation}
Each pair is `sitting' on a different factor (value of $m$) and so the average in the state $\rho_{\r,M}$ of the pair \eqref{47} is the product of the single-site averages,
\begin{equation}
\omega_\r\Big(w_{\pi(1)} w_{\pi(2)}\Big) \cdots \omega_\r\Big(w_{\pi(n-1)} w_{\pi(n)}\Big).
\label{48}
\end{equation}
As discussed above, each term \eqref{48} has a multiplicity $M(M-1)\cdots(M-k_2+1)$ with $k_2=n/2$, which is compensated by the prefactor $M^{-n/2}$,
\begin{equation}
\lim_{M\rightarrow\infty} \frac{M(M-1)\cdots(M-n/2+1)}{M^{n/2}} =1.
\end{equation}
The formula \eqref{45} follows. This completes the proof of Proposition \ref{prop2}.\hfill $\square$

\subsection{Equivalent Gaussian reservoir: Proof of the theorem}
\label{sec:ebgr}

Let $\rho_\f$ be a Gaussian state of a number $N$ (to be determined) of independent quantum harmonic oscillators with creation and annihilation operators $a^\dag_j$, $a_j$, $j=1,\ldots,N$. For $\z=(z_1,\ldots,z_N) \in\mathbb C^N$ we set 
$$
\Phi_\f(\z)=\frac{1}{\sqrt 2}\big(a^\dag(\z)+a(\z)\big),\qquad a^\dag(\z) = \sum_{j=1}^N z_j a_j^\dag, \quad a(\z) = \sum_{j=1}^N \overline z_j a_j.
$$
Here, $\bar z$ is the complex conjugate of $z\in\cx$. Wick's theorem gives
\begin{align}
{\rm tr}_\f\Big(\rho_\f\, \Phi_\f(
\z_1)\cdots \Phi_\f(\z_n)\Big) = \left\{
\begin{array}{cl}
\displaystyle \sum_{\pi\in\mathcal P_n} \prod_{j=1}^{n/2} {\rm tr}_\f\Big(\rho_\f\,  \Phi_\f(\z_{\pi(2j-1)}) \Phi_\f(\z_{\pi(2j)})\Big) & \mbox{for $n$ even} \\ 
\displaystyle 0 & \mbox{for $n$ odd}
\end{array}
\right. 
\label{45-1}
\end{align}
where $\mathcal P_n$ is defined after \eqref{45}. Consider a (`Bogoliubov') dynamics generated by a Hamiltonian $H_\f$ such that
\begin{equation}
	e^{itH_\f} \Phi_\f(\z) e^{-it H_\f} = \Phi_\f(\z(t)),
\end{equation}
where $\z(t)\in\cx^N$ for all $t$. We let the system $\s$ interact with the bosonic reservoir according to the total Hamiltonian,
\begin{equation}
\label{m61}
H_{\s\f} = H_\s +H_\f +\sum_{q=1}^Q G_q\otimes X_q + G_q^\dag\otimes X_q^\dag
\end{equation}
where we take $X_q$ to be 
\begin{equation}
\label{X}
X_q = a(\z_q) + a^\dag(\w_q) = \sum_{j=1}^N \overline z_{q,j} a_j + w_{q,j} a^\dag_j,
\end{equation}
for some $\z_q=(z_{q,1},\ldots,z_{q,N})$ and  $\w_q=(w_{q,1},\ldots,w_{q,N})\in\mathbb C^N$ to be determined.
\medskip

Our strategy to prove the theorem is this: We show that the Dyson series obtained for ${\rm tr}_\f (e^{itH_\s}e^{-it H_{\s\f}} (\rho_\s\otimes\rho_\f) e^{it H_{\s\f}} e^{-itH_\s})$ (see \eqref{m65} below) is the same as the Dyson series \eqref{35.1} in the limit $M\rightarrow\infty$. For the latter Dyson series, we use Proposition \ref{prop2}. Namely, as explained in the proof of Proposition \ref{prop1}, expanding the multiple commutator in the series in \eqref{35.1} gives an operator shown on the right hand side of \eqref{16.1}, and using \eqref{45} we obtain for this term, 
\begin{align}
\lim_{M\rightarrow\infty} &{\rm tr}_\r  \Big( V^{\sigma_{\ell_1}}_{\r,M,r_{\ell_1}}(t_{\ell_1})\cdots V^{\sigma_{\ell_L}}_{\r,M,r_{\ell_L}}(t_{\ell_L})\ \rho_{\r,M}\  V^{\sigma_{\ell_{L+1}}}_{\r,M,
r_{\ell_{L+1}}}(t_{\ell_L+1})\cdots V^{\sigma_{\ell_n}}_{\r,M,r_{\ell_n}}(t_{\ell_n})\Big)\nonumber\\
& = \lim_{M\rightarrow\infty} {\rm tr}_\r  \Big( \rho_{\r,M}\  W_{\ell_{L+1}}\cdots W_{\ell_n} W_{\ell_1}\cdots W_{\ell_L} \Big)\nonumber\\
 &  =
 \left\{
 \begin{array}{cl}
 	\displaystyle \sum_{\pi\in\mathcal P_n} \prod_{j=1}^{n/2} \omega_\r\Big( w_{\pi(2j-1)} w_{\pi(2j)}\Big) & \mbox{for $n$ even}\\
 	\displaystyle 0 & \mbox{for $n$ odd}
 \end{array}
\right.
\label{m89}
\end{align}
where the notation is as in Proposition \ref{prop2} and the  $w_j$ are operators $v_r^\sigma(t)$. 

We now consider the Dyson series resulting from the coupling of $\s$ to the fluctuation reservoir $\f$ with total Hamiltonian $H_{\s\f}$, \eqref{m61}. We write for the free evolution of $X_q$,
\begin{equation}
\label{X(t)}
X_q(t) = e^{i t H_\f} X_q   e^{-i t H_\f} = \sum_{j=1}^N \overline z_{q,j}  e^{-it\omega_j} a_j + w_{q,j} e^{i\omega_j t}a^\dag_j.
\end{equation}
Analogously to \eqref{13-1-1}, we obtain the Dyson series expansion, 
\begin{align}
{\rm tr}_\f & \Big(e^{itH_\s}e^{-it H_{\s\f}} (\rho_\s\otimes\rho_\f) e^{it H_{\s\f}} e^{-itH_\s}\Big)
=\sum_{n\ge 0}(-i)^n \sum_{q_1,\ldots,q_n} \sum_{\sigma_1,\ldots,\sigma_n} \int_{0\le t_n\le\cdots \le t_1\le t} dt_1\cdots dt_n \nonumber\\
&\times \ {\rm tr}_\f\Big( 
\big[G_{q_1}^{\sigma_1}(t_1)\otimes X_{q_1}^{\sigma_1}(t_1), [ \cdots [G_{q_n}^{\sigma_n}(t_n)\otimes X_{q_n}^{\sigma_n}(t_n),\rho_\s\otimes\rho_\f]\ldots ]\big]\Big).
\label{m65}
\end{align}
Consider fixed values of $q_1,\ldots,q_n$ and $\sigma_1,\ldots,\sigma_n$. Expanding the multiple commutator in \eqref{m65} yields $2^n$ terms, each of the form
\begin{align}
	G_{q_{\ell_1}}^{\sigma_{\ell_1}}(t_{\ell_1})\cdots & \, G_{q_{\ell_L}}^{\sigma_{\ell_L}}({t_{\ell_L}}) \ \rho_\s \ G_{q_{\ell_{L+1}}}^{\sigma_{\ell_{L+1}}}(t_{\ell_L+1})\cdots G_{q_{\ell_L}}^{\sigma_{\ell_n}}(t_{\ell_n})\nonumber\\
	&\otimes X_{q_{\ell_1}}^{\sigma_{\ell_1}}(t_{\ell_1})\cdots X_{q_{\ell_L}}^{\sigma_{\ell_L}}(t_{\ell_L})\ \rho_\f\  X_{q_{\ell_{L+1}}}^{\sigma_{\ell_{L+1}}}(t_{\ell_L+1})\cdots X_{q_{\ell_L}}^{\sigma_{\ell_n}}(t_{\ell_n}),
	\label{16.1-1}
\end{align}
just as in \eqref{16.1}. The trace over the reservoir operators in \eqref{16.1-1} is
\begin{align}
{\rm tr}_\f & \Big( X_{q_{\ell_1}}^{\sigma_{\ell_1}}(t_{\ell_1})\cdots X_{q_{\ell_L}}^{\sigma_{\ell_L}}(t_{\ell_L})\ \rho_\f\  X_{q_{\ell_{L+1}}}^{\sigma_{\ell_{L+1}}}(t_{\ell_L+1})\cdots X_{q_{\ell_L}}^{\sigma_{\ell_n}}(t_{\ell_n})\Big)\nonumber\\
& = {\rm tr}_\f \Big(\rho_\f\  X_{q_{\ell_{L+1}}}^{\sigma_{\ell_{L+1}}}(t_{\ell_L+1})\cdots X_{q_{\ell_L}}^{\sigma_{\ell_n}}(t_{\ell_n}) X_{q_{\ell_1}}^{\sigma_{\ell_1}}(t_{\ell_1})\cdots X_{q_{\ell_L}}^{\sigma_{\ell_L}}(t_{\ell_L})\Big)\nonumber\\
& = \left\{
\begin{array}{cl}
	\displaystyle \sum_{\pi\in\mathcal P_n} \prod_{j=1}^{n/2} \omega_\f \Big( x_{\pi(2j-1)} x_{\pi(2j)}\Big) & \mbox{for $n$ even}\\
	\displaystyle 0 & \mbox{for $n$ odd}
\end{array}
\right.
\label{m93}
\end{align}
Here  we set $x_j$ to stand for the $X_q^\sigma(t)$, analogously to the $w_j$ in \eqref{m89}. In the last equality in \eqref{m93} we used Wick's theorem which holds for any linear combination of field operators (hence for the $X_q^\sigma(t)$), see the section \ref{sec:wickthm} in the supplemental material.

Comparing \eqref{m89} and \eqref{m93} we conclude that
$$
 {\rm tr}_\f  \Big(e^{itH_\s}e^{-it H_{\s\f}} (\rho_\s\otimes\rho_\f) e^{it H_{\s\f}} e^{-itH_\s}\Big)
= \lim_{M\rightarrow\infty} {\rm tr}_{\r,M}  \Big( e^{it H_M^0} e^{-it H_{\s\r,M}} \rho_{\s\r,M}\, e^{it H_{\s\r,M}}e^{-it H_M^0}\Big) 
$$
provided that the following two-point functions coincide for each $q, \sigma, t, q', \sigma', t'$,
\begin{equation}
\omega_\r\big( v_q^\sigma(t) v_{q'}^{\sigma'}(t')\big) = \omega_\f\big( X_q^\sigma(t) X_{q'}^{\sigma'}(t')\big).
\label{m94}
\end{equation}
Therefore, the theorem is proven provided we can show \eqref{m94}, which is what we do now.

For a general state $\omega$ and operator $A$ we have $\omega(A)=\overline{\omega(A^\dag)}$. Using this and that $v_q^\dag(t)=(v_q(t))^\dag$ and $X_q^\dag(t)=(X_q(t))^\dag$ shows that \eqref{m94} is equivalent to 
\begin{eqnarray}
\omega_\r\big( v_q(t) v_{q'}(t')\big) &=& \omega_\f\big( X_q(t) X_{q'}(t')\big)\label{c1}\\
\omega_\r\big( v_q^\dag(t) v_{q'}(t')\big) &=& \omega_\f\big( X_q^\dag(t) X_{q'}(t')\big)\label{c2}\\
\omega_\r\big( v_q(t) v_{q'}^\dag(t')\big) &=& \omega_\f\big( X_q(t) X_{q'}^\dag(t')\big)\label{c3}.
\end{eqnarray}
We now identify $\z_q,\w_q$ in \eqref{X} such that \eqref{c1}-\eqref{c3} hold. Denote the Bohr energies of $h_\r$ by $E_{kl}=E_k-E_l$  
and write $[v_q]_{kl}=\langle\chi_k|v_q|\chi_l\rangle$ for the matrix elements of $v_q$ in the eigenbasis 
 $\{\chi_j\}$ of $h_\r$ (see \eqref{6.2}). Using \eqref{6.2} we obtain
\begin{eqnarray}
\omega_\r\big( v_q(t) v_{q'}(t')\big) &=& \sum_{k,l} p_k e^{i(t-t')E_{kl}} [v_q]_{kl} [v_{q'}]_{lk}\label{c1'}\\
\omega_\r\big( v_q^\dag(t) v_{q'}(t')\big) &=& \sum_{k,l} p_k e^{i(t-t')E_{kl}} \overline{[v_q]_{lk}} [v_{q'}]_{lk}\label{c2'}\\
\omega_\r\big( v_q(t) v_{q'}^\dag(t')\big) &=& \sum_{k,l} p_k e^{i(t-t')E_{kl}} [v_q]_{kl}\overline{[v_{q'}]_{kl}}\label{c3'}.
\end{eqnarray}
We can restrict the sums to indices $(k,l)\in\mathcal M$ for which the summands are nonzero, see \eqref{30}.
Next we calculate the right hand sides of \eqref{c1}-\eqref{c3}, with $X_q(t)$ given in \eqref{X(t)}, and where $\omega_\f$ is the product state of the vacua of each mode. We have $\omega_\f(a_ja^\dag_m)=\delta_{j,m}$ (Kronecker) and so
\begin{equation}
\omega_\f\big(X_q(t) X_{q'}(t')\big) = \sum_{j,m} \overline z_{q,j} w_{q',m} e^{-it\omega_j+it'\omega_m }\omega_\f\big(a_j a^\dag_m\big) = \sum_j \overline z_{q,j} w_{q',j} e^{i(t-t')(-\omega_j)}.
\label{c1''}
\end{equation}
We thus take one fluctuation mode for each value of the pair $(k,l)$. The total number of fluctuation modes is (maximally) $N=(\dim\h_\r)^2$ | if $p_k=0$ for some $k$ or some of the matrix elements vanish, then the corresponding terms in the sums \eqref{c1'}-\eqref{c3'} are absent and the number of oscillators is reduced accordingly. Relabeling the fluctuation modes in \eqref{c1''} we write
\begin{equation}
\omega_\f\big(X_q(t) X_{q'}(t')\big) =  \sum_{k,l}  \overline z_{q,kl} w_{q',kl} \, e^{i(t-t')(-\omega_{kl})}.
\label{d}
\end{equation}
In order to have the equality \eqref{c1} we compare \eqref{c1'} and \eqref{d} and identify,
\begin{equation}
z_{q,kl} = \sqrt{p_k}\, \overline{[v_q]_{kl}},\quad w_{q,kl}=\sqrt{p_k} [v_q]_{lk},\quad \omega_{kl} = -E_{kl}=E_{lk}.
\label{m104}
\end{equation}
One then easily checks that the choice \eqref{m104} also guarantees that \eqref{c2} and \eqref{c3} are satisfied. 

This shows that \eqref{m94} holds, and thus the proof of the theorem is complete. \hfill$\square$

\section{Emergence of the ground state for $\f$}
\label{sec:emerGS}

The purpose of this section is to illustrate how the ground state  emerges as the standard representation of $\f$, regardless of the details of the $\s\r$ interaction or the properties of $\r$. We also explain that in certain special cases (of $\s\r$ interactions and properties of $\r$), other representations than the ground state can be used for $\f$. 

To explain the emergence of the ground state of $\f$, we first consider the case \eqref{M} with $Q=1$ and $G=G^\dag$, $v=v^\dag$. As explained in the main text,  we need to find a Gaussian state $\rho_\f$ of $N$ oscillators to match the two-point function \eqref{match5.1}. Denote the `field operator' of $N$ bosonic modes by $\Phi(\z) = \frac{1}{\sqrt 2}(a^\dag(\z)+a(\z))$, $z\in\mathbb C^N$ (compare also with \eqref{X}). A Gaussian state is determined uniquely by its covariance matrix $\mathcal C$ via the expression ${\rm tr}_\f (\rho_\f \,e^{i\Phi(\z)}) = e^{-\frac14\langle \z,\mathcal C\z\rangle}$, where $\langle \z,\w\rangle=\sum_{j=1}^N\overline{z_j} w_j$ is the Euclidean inner product of $\mathbb C^N$ and $\mathcal C$ is an $N\times N$ complex matrix satisfying $\mathcal C\ge \bbbone$. The two-point function is,
\begin{equation}
\label{t1}
{\rm tr} \big(\rho_\f \Phi(\z)\Phi(\w)\big) = \frac14 \langle \z, (\mathcal C+\bbbone)\w\rangle +  \frac14 \langle \w, (\mathcal C-\bbbone)\z\rangle.
\end{equation}
As \eqref{match5.1} depends on the time difference $\tau =t-t'$ only, which encodes the stationarity of $\rho_\r$, the same must be true for \eqref{t1} when $\z$ and $\w$ are replaced by $\z(t)$ and $\w(t')$ (the $j$-th component of $\z(t)$ is $e^{it\omega_j}z_j$). This means that the covariance matrix is diagonal, $\mathcal C={\rm diag}(c_1,\cdots,c_N)$, $c_j\ge 1$, and the time-dependent two-point function \eqref{t1} becomes,
\begin{equation}
\label{t2}
{\rm tr} \big(\rho_\f \Phi(\z(t))\Phi(\w(t'))\big) = \sum_{j=1}^N \Big(\alpha_j \frac{c_j+1}{4} e^{-i\omega_j\tau} + \overline{\alpha}_j \frac{c_j-1}{4} e^{i\omega_j\tau}\Big),\qquad \alpha_j=w_j\overline{z}_j,\quad \tau=t-t'.
\end{equation}
Setting $x_j={\rm Re}\alpha_j$,  $y_j={\rm Im}\alpha_j$ and comparing the real and imaginary parts of \eqref{t2} and \eqref{match5.1} gives,
\begin{align}
\sum_{j=1}^N \frac{c_j}{2} \Big[x_j \cos(\omega_j\tau) + y_j\sin(\omega_j\tau)\Big] &=\sum_{k,l=1}^{d_\r} p_k|v_{kl}|^2 \cos\big((E_l-E_k)\tau\big),\label{t3} \\
\sum_{j=1}^N \frac12 \Big[-x_j \sin(\omega_j\tau)+y_j\cos(\omega_j\tau)\Big] &=\sum_{k,l=1}^{d_\r} -p_k|v_{kl}|^2 \sin\big((E_l-E_k)\tau\big).
\label{t4}
\end{align}
In order to satisfy \eqref{t3} and \eqref{t4} we can take one oscillator $j$ for each ordered pair $(k,l)$, so that $x_j\rightarrow x_{kl}$ and so on.  
Then $\omega_{kl}=E_l-E_k$, $y_{kl}=0$, $ \frac{c_{kl}}{2}x_{kl} = p_k |v_{kl}|^2 = \frac12 x_{kl} $ and therefore $c_{kl}=1$, which means that $\mathcal C=\bbbone$ and $\rho_\f$ is the vacuum state. The same analysis shows that $c_j=1$ in the general case ($Q>1$, $G_q$ and $v_q$ not hermitian). 
\smallskip

{\em Other representations for special cases.} Assume that $Q=1$ in the interaction \eqref{M} and $G=G^\dag$, $v=v^\dag$, with $v_{kk}=0$. Assume also that all the populations $p_j$ are distinct. By a possible relabeling of the energies $E_1,\ldots,E_{d_\r}$ we achieve that 
$$
p_1<p_2<\cdots <p_{d_\r}.
$$
We now show that we can satisfy \eqref{t3} and \eqref{t4} by associating to each pair $(k,l)$ with $k<l$ a mode $a_{kl}$ in a thermal state, accounting for transitions $E_k\rightarrow E_l$ and $E_l\rightarrow E_k$ jointly. Namely, the right hand sides of \eqref{t3} and \eqref{t4} can be written respectively as (use $v_{kk}=0$ and $|v_{kl}|=|v_{lk}|$)
\begin{equation}
\sum_{k<l} (p_k+p_l)|v_{kl}|^2 \cos\big((E_k-E_l)\tau\big)\qquad \mbox{and}\qquad 
\sum_{k<l} (p_k-p_l)|v_{kl}|^2 \sin\big((E_k-E_l)\tau\big).
\label{t5}
\end{equation}
We identify (choose modes $j$ with) $j\leftrightarrow \{(k,l) \mbox{\ with\ } k<l\}$ and we have  
$$
y_{kl}=0,\quad  \omega_{kl}=E_l-E_k,\quad  c_{kl}x_{kl} = 2(p_l+p_k)\, |v_{kl}|^2, \quad x_{kl} = 2(p_l-p_k)\, |v_{kl}|^2. 
$$
The resulting covariance is $c_{kl} =(p_l+p_k)/(p_l-p_k)$ (which is $\ge 1$ as it has to be). This covariance defines the state 
$$
\rho_{kl} = Z^{-1}_{kl}\  e^{-\beta_{kl}\omega_{kl}a^\dag_{kl}a_{kl}},\qquad \text{with} \qquad \beta_{kl} \omega_{kl} = \ln\big(\frac{p_l}{p_k} \big).
$$
With this identification of the fluctuation modes, the reservoir state is $\rho_\f = \otimes_{k<l} \,  \rho_{kl}$, the product of single mode equilibrium states at different temperatures. However, contrary to the standard representation of $\f$ by the ground state, this `thermal representation' cannot be done in general, as shows the following simple example.  

Let $\s$ be arbitrary and let the components of $\r$ be two-level systems with basis states $\{|1\rangle,|2\rangle\}$ and $h_\r=E_1|1\rangle\langle 1|+E_2|2\rangle\langle 2|$, some $E_1,E_2\in\mathbb R$. The the reservoir initial state $\rho_\r=p|1\rangle\langle 1|+(1-p)|2\rangle\langle 2|$ for some $p\in [0,1]$ and take the interaction \eqref{M} with $Q=1$, $G$ arbitrary and $v=v_{12}|1\rangle\langle 2|+v_{21}|2\rangle\langle 1|$ off-diagonal, some $v_{12}, v_{21}\in\mathbb C$ ($v$ not necessarily hermitian). The reservoir $\r$ two-point function (with $\tau=t-t'$, $E_{12}=E_1-E_2$),
\begin{align}
&\mathrm{tr} \big( \rho_\r v(t) v(t') \big) =   p v_{12}v_{21}  e^{i\tau E_{12}} + (1-p) v_{12}v_{21} e^{-i\tau E_{12}} \label{2point_function_vv'}
\end{align}
has to be matched by the reservoir $\f$ two-point correlation function of a Gaussian $\rho_\f$ with covariance $\mathcal C$ (see before \eqref{t1}), which for  $X(\z,\w) = \frac{1}{\sqrt{2}}(a^\dagger (\z) +a(\w) )$, $\z,\w \in \mathbb{C}^N$, is given by (similarly to \eqref{t2}, with $\z(t)$ having components $e^{i\omega_j t}z_j$)
\begin{align}
&\mathrm{tr} \Big(\rho_\f  X\big( \z(t) ,\w(t)\big) X\big(\z(t'),\w(t')\big) \Big) = \sum_{j=1}^N \Big( \frac{c_j+1}{4} \bar{w}_j z_j e^{-i\tau\omega_j} + \frac{c_j-1}{4} \bar{z}_j w_j e^{i\tau\omega_j} \Big) \label{2point_function_XX'}.
\end{align}
Suppose we could represent $\f$ by a single oscillator (in a thermal state, as outlined above), then $N=1$ in \eqref{2point_function_XX'}. 
Equating \eqref{2point_function_vv'} and \eqref{2point_function_XX'} implies that $\omega$ is either $E_{12}$ or $-E_{12}$. If $\omega=-E_{12}$ then $(c+1)\bar w z=4pv_{12}v_{21}$ and  $(c-1)\bar z w=4(1-p) v_{12}v_{21}$. Take $p\neq 0,1$. The last two equalities imply that $v_{12}v_{21} =\frac{1}{4p}(c+1)\bar wz=\frac{1}{4(1-p)}(c-1)\bar z w$. Therefore, as $c\ge 1$ and $p$ are real,  we must have $\bar z w\in\mathbb R$. This forces $v_{12}v_{21}\in\mathbb R$. The analysis for $\omega=E_{12}$ yields the same result. Consequently, a simple case where we cannot realize $\f$ by a thermal mode is when the phases of $v_{12}$ and $v_{21}$ are not opposite.

\section{Wick's theorem}
\label{sec:wickthm}
\setcounter{equation}{0}

Let $\h$ be a Hilbert space and let $W(f)$ be Weyl operators on $\h$, for $f\in\hh$ and where $\hh$ is another Hilbert space, the so-called single-particle Hilbert space. (For $N$ bosonic modes we have $\hh=\mathbb C^N$, for a `usual' scalar field in three space dimensions, $\hh=L^2(\mathbb R^3,d^3x)$.) The operators $W(f)$ are unitary and satisfy
\begin{equation}
\label{a1}
W(-f)=W(f)^*\qquad \mbox{and}\qquad W(f)W(g) = e^{-\frac i2{\rm Im }\langle f,g\rangle} W(f+g),
\end{equation}
where $\langle\cdot,\cdot\rangle$ is the inner product of $\hh$. We assume that 
$$
W(f) = e^{i\Phi(f)}
$$
for some self-adjoint operator $\Phi(f)$ on $\h$, which is called the field operator  (the field operators exist for so-called `regular representation' of the canonical commutation relations \cite{petz}). $\Phi$ is real linear, that is, $\Phi(tf)=t\Phi(f)$, $t\in\rx$. Let $\rho$ be a density matrix on $\h$ satisfying 
\begin{equation}
\label{a3}
{\rm tr}\big(\rho W(f)\big) = e^{-\frac14\langle f, \mathcal C f\rangle},
\end{equation}
where $\mathcal C\ge \bbbone $ is an operator called the covariance operator. The state $\rho$ is called centered and Gaussian (or, quasi-free). We want to find an expression for the correlation function ${\rm tr}\big(\rho \Phi(f_1)\cdots \Phi(f_n)\big)$, where $n\ge 1$ and $f_j\in\hh$ are given. As $\Phi(f)=(-i)\partial_t|_0 W(tf)$ and due to \eqref{a1}, \eqref{a3} we have 
$$
{\rm tr}\big(\rho \Phi(f_1)\cdots \Phi(f_n)\big) = (-i)^n\partial^n_{t_1,\ldots,t_n}\big|_0 \ e^{-\frac i2\sum_{k<l}t_kt_l{\rm Im }\langle f_k,f_l\rangle} e^{-\frac14 \sum_{k,l}t_kt_l\langle f_k,\mathcal Cf_l\rangle}.
$$
The term in the second exponent when $k=l$ yields squares $t_k^2$ which do not contribute as we take the derivatives and set all $t_k$ equal to zero in the end. Therefore, 
$$
{\rm tr}\big(\rho \Phi(f_1)\cdots \Phi(f_n)\big) = (-i)^n\partial^n_{t_1,\ldots,t_n}\big|_0 \ e^{\sum_{k<l} t_kt_l q_{kl}},\quad 
q_{kl} = -\tfrac12\big( {\rm Re}\langle f_k,\mathcal Cf_l\rangle + i {\rm Im }\langle f_k,f_l\rangle\big).
$$
Next,
$$
\partial^n_{t_1,\ldots,t_n}\big|_0 e^{\sum_{k<l} t_kt_l q_{kl}} = \partial^n_{t_1,\ldots,t_n}\big|_0 \sum_{r\ge 0} \frac{1}{r!}\Big(\sum_{k<l} t_kt_l q_{kl}\Big)^r 
$$
is nonzero only if $n$ is even and $r=n/2$, because the $t$ appear in pairs. Hence
\begin{equation}
\label{a4}
{\rm tr}\big(\rho \Phi(f_1)\cdots \Phi(f_n)\big) = 
 (-1)^{n/2}\partial^n_{t_1,\ldots,t_n}\big|_0  \frac{1}{(n/2)!}\Big(\sum_{k<l} t_kt_l q_{kl}\Big)^{n/2}.
\end{equation}
The product of the sums leads to polynomials in the $t$ and only those terms in which the polynomial is $t_1\cdots t_n$ are nonzero after taking the derivative. We must thus choose in each sum (factor) one pair $k<l$ in such a way that all pairs make up the indices $\{1,\ldots,n\}$. The associated value to each choice is 
\begin{equation}
\label{a5}
q_{k_1,l_1}\cdots q_{k_{n/2}l_{n/2}}.
\end{equation}
As the order in which we choose the pairs does not matter for the resulting (`commutative') value \eqref{a5}, we get a multiplicity $(n/2)!$ for each such value. This removes the prefactor $1/{(n/2)!}$ in \eqref{a4}. We may list the factors such that $k_1<k_2<\cdots$ and the order $k_j<l_j$ is imposed by \eqref{a4}. Summing over all such arrangements thus yields the value of \eqref{a4}. Finally, we note that $-q_{kl} = {\rm tr}\big(\rho\, \Phi(f_k)\Phi_f(l)\big)$. 
We have derived {\it Wick's theorem}, stating that 
\begin{equation}
\label{a6}
{\rm tr}\big(\rho \Phi(f_1)\cdots \Phi(f_n)\big) = \left\{
\begin{array}{cl}
\displaystyle 0 & \mbox{for $n$ odd}\\
\displaystyle \sum_{\pi\in\mathcal P_n} \prod_{j=1}^{n/2} {\rm tr}\Big(\rho\,  \Phi(f_{\pi(2j-1)}) \Phi(f_{\pi(2j)})\Big) & \mbox{for $n$ even}
\end{array}
\right. 
\end{equation}
where $\mathcal P_n$ is the set of permutations $\pi$ satisfying \eqref{perm'}.

We note that if in \eqref{a1} the commutation relation is replaced by $W(f)W(g)=W(f+g)$, then the above derivation and the result \eqref{a6} hold in the exact same way. This corresponds to a commutative, or classical representation of the canonical commutation relations. 

It is sometimes useful to state Wick's theorem for operators more general than field operators (and indeed, this is what we do in the proof of the theorem in the letter). Since both sides in \eqref{a6} are linear in each $\Phi(f_j)$ we may replace each of those field operators by any linear combination,
$$
\Phi(f_j)\mapsto X_j\equiv \sum_{r=1}^R \xi_{r,j}\Phi(g_{r,j}),
$$
for any $\xi_{r,j}\in\mathbb C$, 
and the relation \eqref{a6} stays preserved,
\begin{equation}
\label{a6'}
{\rm tr}\big(\rho X_1\cdots X_n\big) = \left\{
\begin{array}{cl}
\displaystyle 0 & \mbox{for $n$ odd}\\
\displaystyle \sum_{\pi\in\mathcal P_n} \prod_{j=1}^{n/2} {\rm tr}\Big(\rho\,  X_{\pi(2j-1)} X_{\pi(2j)} \Big) & \mbox{for $n$ even}
\end{array}
\right. 
\end{equation}
A particular example are creation and annihilation operators,
$$
a^\dag(f) = \frac{1}{\sqrt 2}\Big(\Phi(f)-i\Phi(if)\Big),\qquad a(f) = \frac{1}{\sqrt 2}\Big(\Phi(f)+i\Phi(if)\Big),
$$
for which we have
\begin{equation}
\label{a7}
{\rm tr}\big(\rho a^{\sigma_1}(f_1)\cdots a^{\sigma_n}(f_n)\big) = \left\{
\begin{array}{cl}
\displaystyle 0 & \mbox{for $n$ odd}\\
\displaystyle \sum_{\pi\in\mathcal P_n} \prod_{j=1}^{n/2} {\rm tr}\Big(\rho\,  a^{\sigma_{\pi(2j-1)}}(f_{\pi(2j-1)}) a^{\sigma_{\pi(2j)}}(f_{\pi(2j)})\Big) & \mbox{for $n$ even}
\end{array}
\right. 
\end{equation}
where $\sigma_j\in\{1,-1\}$ and
$$
a^\sigma = 
\left\{\begin{array}{ll}
a, & \sigma=1\\
a^\dag,& \sigma=-1
\end{array}
\right. 
.
$$

\end{document}